\documentclass[%
reprint,
amsmath,amssymb,
aps, superscriptaddress
pra
]{revtex4-2}

\usepackage{graphicx}
\usepackage{dcolumn}
\usepackage{bm}
\usepackage{amsmath}
\usepackage{mathrsfs}
\usepackage{mathtools}
\usepackage{xcolor}
\usepackage{orcidlink}

\begin{document}

\preprint{APS/123-QED}

\title{Emergent time crystal from a fractional Langevin equation \\ with white and colored noise}

\author{D.~S. Quevedo\orcidlink{0000-0003-1583-4262}}
\email{d.s.quevedovega@uu.nl}
 \affiliation{Institute for Theoretical Physics, Utrecht University, Princetonplein 5, 3584CC Utrecht, The Netherlands}
\author{R.~C. Verstraten\orcidlink{0000-0002-4386-5210}}%
\affiliation{Institute for Theoretical Physics, Utrecht University, Princetonplein 5, 3584CC Utrecht, The Netherlands}
\author{C. Morais Smith\orcidlink{0000-0002-4190-3893}}%
\affiliation{Institute for Theoretical Physics, Utrecht University, Princetonplein 5, 3584CC Utrecht, The Netherlands}

\date{\today}

\begin{abstract}
We study the fractional Langevin equation with fractional $\alpha$-order and linear friction terms of a system coupled to white and colored thermal baths using both analytical and numerical methods. We find analytical expressions for the position and the mean squared displacement (MSD) of the system using the Prabhakar-Mittag-Leffler function. The MSD exhibits long-term sub-diffusive regimes $t^\alpha$ driven by colored noise and $t^{2\alpha-1}$ driven by white noise. When the linear friction is neglected, periodic ordered phases emerge for small fractional orders $\alpha \lesssim 0.1$. In particular, the zero-linear friction system driven only by colored noise manifests the properties of a time crystal, with a ground state satisfying the fluctuation-dissipation theorem and a periodicity proportional to $2\pi$. On the other hand, the zero-linear friction system driven only by white noise displays an out-of-equilibrium time glass phase with periodicity proportional to $\pi$. A mixed phase with contributions from both ground and out-of-equilibrium states is encountered when the system couples to both baths. In that case, the periodicity deviates from $2\pi$ due to damping effects. We test all the analytical results numerically by implementing a discrete recursive expression, where the random forces of the system are modeled as the derivative of the fractional Brownian motion. A microscopic description for the system is also provided by an extension of the Caldeira-Leggett model.
\end{abstract}

\maketitle


\section{\label{sec:introduction}Introduction}

Anomalous diffusion manifests in several complex systems~\cite{metzler2000random, metzler2004restaurant}. Conventionally, it can be defined as the deviation of the mean squared displacement (MSD) from the linear behavior in $t$ described as normal diffusion, such that $\langle x^2 \rangle \propto t^{\alpha}$~\cite{hanyga2007anomalous}. When $0 < \alpha < 1$, $x(t)$ is said to exhibit sub-diffusive behavior; conversely, the regime $1 < \alpha < 2$ is called super-diffusive. Experimental observations have shown that anomalous diffusion is triggered by non-local behavior, such as the persistence of memory effects, and recent advances in particle tracking techniques have shown that it occurs in several biophysical systems under the influence of mixtures of noises with different correlation times~\cite{burov2011single, weigel2011ergodic, wang2022anomalous}. Furthermore, in many cases these effects also arise in multiple spatio-temporal scales. As a consequence, the probability density functions lack the characteristic spatio-temporal scaling in displacement~\cite{tateishi2017role}.

Several models have been formulated in the past decades for the description of anomalous diffusion. A very interesting case is when the generalized and fractional Langevin equations (fLe) describe systems coupled to different types of thermal baths~\cite{porra1996generalized, lutz2001fractional, fa2006generalized, vinales2007anomalous, tateishi2012different, sandev2014langevin}. In Ref.~\cite{tateishi2012different}, the generalized Langevin equation for a free particle coupled to white- and colored-noise thermal baths was studied, showing the existence of an asymptotic linear behavior in the MSD for the short-term and super- and sub-diffusive regimes in the long-term. This idea was extended in Ref.~\cite{sandev2014langevin} to a mixture of $n$ power-law correlated noises and revisited in Ref.~\cite{geraghty2019financial} to characterize the financial market using a fLe solved via Monte Carlo simulations. So far, the asymptotic limits of these Langevin-type equations have been thoroughly investigated. In particular, in the presence of friction terms with kernel functions of the type $t^{-\alpha}$, it is well known that the diffusive coefficient behaves as $t^{\alpha}$ and the MSD presents sub- and super-diffusive regimes in the long term~\cite{morgado2002relation}. However, the emergence of diffusive regimes that exhibit time-ordered phases has received much less attention.

The idea of phases of matter with spontaneous periodic temporal ordering was introduced by Frank Wilczek in 2012 in the context of time crystals, drawing an analogy to the spatial ordering found in regular crystals~\cite{shapere2012classical,wilczek2012quantum}. After the formulation of no-go theorems for systems in equilibrium~\cite{bruno2012comment, watanabe2015absence}, it was understood that time crystals could arise in open quantum systems~\cite{khemani2016phase}. This concept has ignited intense research utilizing both analytical and experimental techniques~\cite{sacha2017time,sacha2020time,else2020discrete,zhang2017observation,kyprianidis2021observation,kongkhambut2022observation,kessler2021observation}. Subsequently, Wilczek conjectured the existence of other ``time materials"~\cite{wilczek2019crystals}, which have been unveiled in the past years, such as imaginary-time crystals~\cite{cai2020imaginary,arouca2022non}, time quasi-crystals~\cite{giergiel2019discrete}, time liquids and time glasses~\cite{verstraten2021time}. Time glasses are systems exhibiting a glass-like plateau at regular intervals within a larger glassy phase~\cite{elizondo2019glass,khemani2019brief}. In Ref.~\cite{verstraten2021time}, a fLe with fractional friction of order $0\!<\!\alpha\!<\!1$ and white noise was used to study the emergence of a time glass from a mean-field description grounded in the semi-classical limit of the Caldeira-Leggett model for open quantum systems~\cite{caldeira1983path, caldeira1985influence}. It was shown that for sufficiently small values of the fractional order ($\alpha\lesssim 0.1$), damped oscillations with periodicity proportional to $\pi$ emerge, manifesting the clear signature of spontaneous time-symmetry breaking associated with time glasses. More recently, a time rondeau crystal, which exhibits short-term disorder and long-term order was proposed in a system with random drivings \cite{moon2024experimental}.

Here, we revisit the problem of the fLe with linear and colored friction describing a particle coupled to white- and colored-noise thermal baths to investigate the emergence of diffusive regimes with periodic time-ordered phases exhibiting time crystal signatures. All the results are tested by comparing analytical and numerical solutions to provide reliable conclusions. The dissipative component of the system is defined by the superposition of a linear term and a fractional derivative term of order $0\!<\!\alpha\!<\!1$ with damping constants $\gamma$ and $\zeta$, respectively. In the presence of both noises and both dissipative terms, the MSD exhibits the sub-diffusive regimes obtained in Refs.~\cite{tateishi2012different,sandev2014langevin}. However, when the linear friction is removed, oscillatory regimes emerge in the MSD for $\alpha\lesssim 0.1$. This behavior was exclusively linked to the presence of white noise in Ref.~\cite{verstraten2021time}, but our results show similar properties, which can be associated with a time crystal phase for a system with only colored noise and a superposition of time crystal and time glass phases for the mixture of colored and white noises.
 
 This work is structured as follows. A model for the fLe with white and colored noise is introduced in Sec.~\ref{sec:fLe}, together with the main macroscopic implications of the Caldeira-Leggett formulation of the system, and the different regimes that preserve and break fluctuation-dissipation theorem. The description of the numerical scheme obtained by deriving a recursive expression following Ref.~\cite{guo2013numerics} and modeling the noises using fractional Brownian motion is presented in Sec.~\ref{sec:fLe-num}. In Sec.~\ref{sec:fLe-sol}, the system is analytically solved in terms of the Prabhakar-Mittag-Leffler function~\cite{garra2018prabhakar}, and asymptotic limits are provided for the short- and long-time scales when the linear friction is neglected. A detailed comparison between the analytical and numerical results is presented in Sec.~\ref{sec:results}, together with a discussion on the emergent time crystal phases. The formal derivation of the fLe from the microscopic perspective of the Caldeira-Leggett model is presented in Sec.~\ref{sec:derivation}. Conclusions and outlook are presented in Sec.~\ref{sec:con}.

\section{\label{sec:fLe} The fLe for white and colored type thermal baths}

We start by introducing the fLe that describes the motion of a particle of mass $M$ with dissipation given by the superposition of linear and fractional-order friction terms, driven by the action of two independent random forces $f_1(t)$ and $f_2(t)$,  
\begin{equation}
    M\frac{d^2 x(t)}{dt^2} + \gamma \frac{dx}{dt} + \zeta \prescript{C}{0}{D}^{\alpha}_t x(t) = f_1(t) + f_2(t),
    \label{eq:fLe1}
\end{equation}
where $0 < \alpha < 1$, $\gamma$ and $\zeta$ denote general damping coefficients that depend on the equilibrium state of the system, and $\prescript{C}{0}{D}^{\alpha}_t$ stands for the Caputo $(C)$ fractional derivative of order $\alpha$. A brief review about fractional derivatives is given in the Appendix \ref{AppFC}.

The random forces in Eq.~\eqref{eq:fLe1} are stationary second-order processes, that is, the time correlations between two different times $t$ and $\tau$ are functions of the time difference $|t - \tau|$. In particular, we assume that both random forces are associated to two independent thermal baths, such that $f_1(t)$ is white noise with expected value $\langle f_{1}(t) \rangle = 0$ and time correlations 
\begin{equation}
 C_{1}(|t-\tau|) = \langle f_{1}(t) f_{1}(\tau) \rangle = \theta_{1}^2\delta(t-\tau),
 \label{eq:corr_f1}
\end{equation}
and $f_2$ is colored noise with expected value $\langle f_2(t)\rangle = 0$ and correlations 
\begin{equation}
 C_2(|t-\tau|) = \langle f_2(t) f_2(\tau)\rangle = \theta_2^2 \frac{|t-\tau|^{-\alpha}}{\Gamma(1-\alpha)},
 \label{eq:corr_f2}
\end{equation}
where the coefficients $\theta_{1}$ and $\theta_2$ are the strength of the noises. 

The fLe~\eqref{eq:fLe1}, together with Eqs.~\eqref{eq:corr_f1} and \eqref{eq:corr_f2} are obtained from a microscopic formalism by extending the Caldeira-Leggett model, which describes a particle coupled to an environment composed of a large set of harmonic oscillators~\cite{caldeira1983path, caldeira1985influence}. We present a detailed derivation of these equations in Sec.\ref{sec:derivation}. Here, we focus on the main macroscopic physical features of the system. For this purpose, we assume an environment consisting of two separate baths, where each bath contributes with one dissipation and one fluctuation term. The connection between the microscopic model and the macroscopic dynamics given by Eq.~\eqref{eq:fLe1} is made using the spectral function $J_i(\omega) \propto \eta_i\omega^{\alpha_i}$ that describes the strength at which the different frequencies of the harmonic oscillators couple to the particle, where $\alpha_i>0$, $\eta_i$ is the damping constant associated to each bath and the sub-indices $i=1,2$ represent the two baths. 

When the system is in equilibrium and the spectral functions of each bath are assumed to be given by the Ohmic and non-Ohmic functions $J_1(\omega)=\eta_1 \omega$ and $J_2(\omega)=\eta_2 \sin(\pi \alpha/2)\omega^\alpha$, the baths exhibit the statistical properties of white and colored noise given by Eqs.~\eqref{eq:corr_f1} and \eqref{eq:corr_f2}, respectively. Furthermore, the Ohmic bath is responsible for the emergence of the linear friction term, while the non-Ohmic bath causes the fractional-friction term. Thus, assuming the equilibrium temperatures of each thermal bath to be $T_1$ and $T_2$, the relations for the strength of the noises are
\begin{align}
  \begin{split}
  \theta_1^2 = 2\eta_1 k_BT_1, \\
  \theta_2^2 = \eta_2 k_BT_2,
  \label{eq:theta_eq}
  \end{split}
\end{align}
and the damping terms are
\begin{align}
  \begin{split}
    \gamma = \eta_1, \\
    \zeta = \eta_2.
  \label{eq:damping_eq}
  \end{split}
\end{align}

Due to the absence of coupling effects between the thermal baths, the cross-time correlations are $\langle f_1(t) f_2(\tau)  \rangle = 0 $. Therefore, the total random force $f(t) \equiv f_1(t) + f_2(t)$ is also a second-order stationary process with zero mean $\left<f(t)\right> = 0$ and time correlations given by the superposition of the correlations of each bath,
\begin{equation}
    \begin{split}
    C(|t-\tau|) & = \langle f(t) f(\tau)\rangle = C_1(|t-\tau|) + C_2(|t-\tau|).
   \label{eq:corr_ft}
   \end{split}
\end{equation}

Using the definition of the Caputo derivative and the Dirac delta function, we express the damping terms in Eq.~\eqref{eq:fLe1} as
\begin{align}
    \begin{split}
    \gamma\frac{dx}{dt} & +  \zeta\prescript{C}{0}{D}^{\alpha}_t x(t) = \\
    & \int_0^t 2\gamma \delta(t-\tau)\dot{x}(\tau)d\tau + \int_0^t \zeta \frac{(t-\tau)^{-\alpha}}{\Gamma(1-\alpha)}\dot{x}(\tau)d\tau,
    \label{eq:damping}
    \end{split}
\end{align}
where $\dot{x} \equiv dx/d\tau$. Furthermore, from Eq.~\eqref{eq:damping}, we identify the memory kernel
\begin{equation}
    \eta(t) = 2 \gamma \delta(t) + \zeta \frac{t^{-\alpha}}{\Gamma(1-\alpha)},
    \label{eq:eta_eq}
\end{equation}
which allows us to rewrite the fLe~\eqref{eq:fLe1} as a conventional generalized Langevin equation
\begin{equation}
   M\frac{d^2 x(t)}{dt^2} + \int_0^t \eta(t-\tau) \dot{x}(\tau) d\tau = f(t).
   \label{eq:gLe}
\end{equation}

Comparing the memory kernel in Eq.~\eqref{eq:eta_eq} with the correlation functions in Eqs.~\eqref{eq:corr_f1},~\eqref{eq:corr_f2},~\eqref{eq:corr_ft}, and setting $T=T_1=T_2$, we observe that the system naturally obeys the second fluctuation-dissipation theorem~\cite{kubo1966fluctuation, porra1996generalized},
\begin{equation}
    \begin{split}
    \left<f(t)f(\tau)\right> = C(|t-\tau|) = k_BT \eta(t - \tau).
    \label{eq:FDT}
    \end{split}
\end{equation}
In this case, the relaxation time of the system and the correlation time of the driving noise are the same, which means that the driving noises are internal.

In the limit $\eta_2 \rightarrow 0$, Eq.~\eqref{eq:fLe1} retrieves the conventional Langevin equation with linear friction and white noise that describes the ordinary Brownian motion. In contrast, the limit $\eta_1 \to 0$ retrieves the fLe with colored noise that, in the long term, exhibits a deviation in the MSD from normal diffusion proportional to $t^{\alpha}$~\cite{lutz2001fractional}. Notice that both limiting cases satisfy the second fluctuation-dissipation theorem and keep the system in equilibrium. Furthermore, they can be pictured from the microscopic description of the system, as a particle coupled to a single bath (with white or colored noise, respectively).

Out-of-equilibrium cases of the fLe~\eqref{eq:fLe1} are observed when the correlation times of the noise and the relaxation times of the system are different. In particular, if $\gamma \to 0$ the correlation time of the noise is given by the superposition \eqref{eq:corr_ft}, while the relaxation time is given by the power-law behavior of the fractional friction. This is naturally obtained from the microscopic description of the system by assuming a local temperature associated with each harmonic oscillator $j$. With this in mind, we introduce the thermal gradient inspired by Ref.~\cite{verstraten2021time}, $T_i\to T_i[t_\alpha \omega_{i,j}]^{1-\alpha}$, where $t_\alpha=(M/\eta_i)^{1/(2-\alpha)}$ and $i=1,2$.

If the thermal gradient is induced in bath $1$ together with the non-Ohmic spectral function $J_1(\omega)= \eta_1 \sin(\pi \alpha/2) \omega^\alpha$, while bath $2$ is kept in thermal equilibrium at $T_2$ by setting $J_2(\omega)= \eta_2 \sin(\pi \alpha/2) \omega^\alpha $, the strength of interaction with the random forces is given by
\begin{align}
  \begin{split}
  \theta_1^2 = 2 \eta_1 k_B\tilde{T}_1 t_\alpha^{1-\alpha}, \\
  \theta_2^2 = \eta_2 k_BT_2,
  \label{eq:theta_neq}
  \end{split}
\end{align}
where $\tilde{T}_1 = T_1 \sin\left( \pi\alpha/2\right)$.

Furthermore, both thermal baths cause fractional friction, and therefore, the damping terms are given by
\begin{align}
  \begin{split}
    \gamma = 0, \\
    \zeta = \eta_1 + \eta_2.
  \label{eq:damping_neq}
  \end{split}
\end{align}

The former situation breaks the second fluctuation-dissipation theorem in a controlled way. This causes a setting in which the particle is coupled to both colored- and white-noise thermal baths, but the friction is exclusively of fractional type. When bath $2$ is kept in thermal equilibrium, there is still a relationship between the memory kernel of the system and the correlation function of bath $2$. By comparing Eqs.~\eqref{eq:corr_f2}, \eqref{eq:FDT}, \eqref{eq:theta_neq}, and \eqref{eq:damping_neq}, we observe
\begin{align}
  \left<f_2(t)f_2(\tau)\right> = C_2(|t-\tau|) = \frac{\eta_2k_BT_2}{\zeta} \tilde{\eta}(t - \tau),
  \label{eq:FDT_weak}
\end{align}
where $\tilde{\eta}(t)$ is the memory kernel in Eq.~\eqref{eq:eta_eq} in the limit $\gamma \to 0$,
\begin{equation}
    \tilde{\eta}(t) = \lim_{\gamma \to 0} \eta(t) = \zeta \frac{t^{-\alpha}}{\Gamma(1-\alpha)} =(\eta_1 + \eta_2) \frac{t^{-\alpha}}{\Gamma(1-\alpha)}.
    \label{eq:eta_neq}
\end{equation}
This relation does not hold if the bath $2$ is neglected, and therefore, $\theta_2=\eta_2=0$. In that case, the relaxation time of the system is power-law type, whereas the white noise is uncorrelated in time.

\section{\label{sec:fLe-num}Numerical solutions of the fLe}

To perform numerical solutions of the fLe \eqref{eq:fLe1}, we model the white and colored noise forces using the derivative of the ordinary and fractional Brownian motions, respectively,
\begin{equation}
    f_1(t) = \mathcal{A}\frac{dB(t)}{dt},
    \label{eq:f1}
\end{equation}
\begin{equation}
    f_2(t) = \mathcal{A}_{H}\frac{dB_H(t)}{dt},
    \label{eq:f2}
\end{equation}
where $\mathcal{A}$ and $\mathcal{A}_{H}$ are normalization constants of the noises, $H$ is the Hurst exponent, such that $0<H<1$, and $B(t) \equiv B_{1/2}(t)$ is ordinary Brownian motion (oBm).

The first definitions of fractional Brownian motion (fBm) were proposed by Kolmogorov in 1940~\cite{kolmogorov1940wienersche} and Lévy in 1953~\cite{levy1953random} using the Riemann-Liouville fractional integral, as a moving average on the increments of an oBm $B(\tau)$, weighted by the function $(t-\tau)^{H-1/2}$. Lately, it was extended and formalized by Mandelbrot and Van Ness in 1968~\cite{mandelbrot1968fractional} by introducing an additional term, based on Weyl's integral, to avoid the ``excessive weight to the location of the origin''. Considering both arguments, the full definition reads
\begin{align}
    \begin{split}
        B_{H} & = \frac{1}{\Gamma(H+1/2)} \biggl\{ \int_{-\infty}^0 [(t-\tau)^{H-1/2} \\
        & - (-\tau)^{H-1/2}] dB(\tau) + \int_0^t (t-\tau)^{H-1/2} dB(\tau) \biggr\}.
    \label{eq:BH_man}
    \end{split}
\end{align}

The Hurst exponent~\cite{hurst1951long} in Eq.~\eqref{eq:BH_man} aims to capture the effect of long-term memory on the trajectories~\cite{friday1981dependence}. When $H > 1/2$ ($H < 1/2$), the increments of the process are positively (negatively) correlated with their history and are said to have a persistent (anti-persistent) long-term dependence. On the other hand, if $H=1/2$ the increments are independent, identically distributed, and the process recovers the oBm.

In the context of time series analysis, $H$ can also be connected to the fractal dimension $D$ of a time series defined on an $n$-dimensional self-affine surface through the relation $D = n + 1 - H$~\cite{mandelbrot1985self, gneiting2004stochastic}. Notice that this expression connects the long-term memory of the stochastic process, which is a global property, to the fractality, which is in general local. Therefore, this is satisfied only by a self-affine process. In particular, for a one-dimensional fBm, we have $H = 2 - D$.

From the definition in Eq.~\eqref{eq:BH_man}, it is possible to verify that the time correlations of the fBm satisfy
\begin{equation}
    \langle B_{H}(t)B_{H}(\tau) \rangle = \frac{\Lambda}{2} \left(t^{2H} + \tau^{2H} + |t-\tau|^{2H}\right),
    \label{eq:corr_fBm}
\end{equation}
and the MSD of the process is $\langle B_{H}(t)^2 \rangle = \Lambda t^{2H}$, which yields the sub-diffusive ($0<H<1/2$), super-diffusive ($1/2<H<1$) and diffusive ($H = 1/2$) limits of the motion. The normalization constant $\Lambda$ can be calculated at $\langle B_{H}(1)^2 \rangle = \Lambda$. This calculation leads to $\Lambda=\Gamma(2-2H)/2H\Gamma(3/2-H)\Gamma(1/2+H)$~\cite{guo2013numerics}. However, for practicality, it is common to set $\Lambda=1$, and we also assume this condition in this work.

By differentiating Eq.~\eqref{eq:corr_fBm}, we see that the time correlation of the colored or fractional Gaussian noise is given by
\begin{equation}
    \langle dB_{H}(t) dB_{H}(\tau) \rangle = H(2H-1)|t-\tau|^{2H-2}dtd\tau,
    \label{eq:corr_fGn}
\end{equation}
where $\langle dB_H(t) \rangle = 0$. Similarly, when $H=1/2$, the derivative of the Brownian trajectory generates white or Gaussian noise $dB(t)/dt$, with mean $\langle dB(t) \rangle = 0$ and correlations
\begin{equation}
    \langle dB(t) dB(\tau) \rangle = \delta(t-\tau)dtd\tau.
    \label{eq:corr_Gn}
\end{equation}

Comparing Eqs.~\eqref{eq:corr_f1} and \eqref{eq:corr_f2} with Eqs.~\eqref{eq:corr_fGn} and \eqref{eq:corr_Gn}, we find that the scaling constants $\mathcal{A}$ and $\mathcal{A}_{H}$ are defined in terms of the Hurst exponent and the coefficients $\theta_1$ and $\theta_2$ characterizing the thermal baths,
\begin{equation}
    \begin{split}
    \mathcal{A} = \theta_1, \hspace{1.5mm} \mathcal{A}_{H} = \frac{\theta_2}{\sqrt{H(2H-1)\Gamma(2H-1)}},
    \label{eq:A_AH}
    \end{split}
\end{equation}
and the fractional-order of the derivative is given by
\begin{equation}
    \alpha = 2 - 2H.
\label{eq:alpha}
\end{equation}

\subsection{\label{sec:fLe-fdiff}Discretization of the fLe}

In Ref.~\cite{geraghty2019financial}, the fLe \eqref{eq:fLe1} was solved using the Monte Carlo method, in the context of a model of the financial market. The method consists of a Next-Guess algorithm that starts from the solution of Eq.~\eqref{eq:fLe1} without the fractional operator, and moves towards the actual solution of the system by minimizing the residual function $Res_k = |h(t_k)- \zeta \prescript{C}{0}{D}^{\alpha}_{t_k} x(t_k)|$, where 
$h(t_k) =  Md^2 x(t_k)/dt_k^2 + \gamma dx_k/dt_k - f(t_k)$
is the integer order part of the equation and $k = {1,2,3,...,N}$ are a set of $N$ points in which the functions are discretized. Due to the non-Markovian characteristics of the system, every time a new point $k$ is guessed, the residuals must be minimized in a large enough set of points $k-1, k-2, ... k-l$ to preserve the memory of the system and the stability of the solution.

This algorithm provides reliable solutions of the fLe, but it is computationally expensive to solve Eq.~\eqref{eq:fLe1} several times in order to have enough realizations to calculate the MSD. Thus, we extend the numerical scheme developed in Ref.~\cite{guo2013numerics} to find a recursive evolution rule for the position $x(t)$, by discretizing the stochastic fractional differential equation~\eqref{eq:fLe1}. This technique has been extensively investigated over the past few years to solve deterministic fractional differential equations in time and space~\cite{li2015numerical}. However, in our case we also work with a linear combination of stochastic terms. 

By replacing Eqs.~\eqref{eq:f1} and \eqref{eq:f2} in Eq.~\eqref{eq:fLe1} and integrating with respect to time, we obtain
\begin{align}
    \begin{split}
        M \left[\dot{x}(t) - v_0 \right] + \gamma \left[ x(t) - x_0 \right] + \zeta \int_0^t \prescript{C}{0}{D}^{2-2H}_{t'}x(t') dt' \\ = \mathcal{A}\left[ B(t) - B(0) \right] +  \mathcal{A}_{H}\left[ B_{H}(t) - B_{H}(0) \right],
    \label{eq:FLE2}
    \end{split}
\end{align}
where $v_0 \equiv \dot{x}(0)$ and $x_0 \equiv x(0)$.

The integral of the Caputo fractional operator can be easily simplified by recalling its relationship with the Riemann-Liouville definition~\eqref{eq:FC_RL} (see Appendix~\ref{AppFC}). For simplicity, we set the initial conditions to zero, both for the position and the (fractional) Brownian trajectories $x_0 = B_{H}(0) = B(0) = 0$. Thus, we find
\begin{equation} 
    \begin{split}
        \dot{x}(t) & + \frac{\gamma}{M} x(t) = v_0 + \frac{\mathcal{A}}{M}B(t) + \frac{\mathcal{A}_{H}}{M}B_{H}(t) \\
         & - \frac{\zeta}{M\Gamma(2H-1)} \int_0^t (t-\tau)^{2H-2} x(\tau) d\tau.
    \label{eq:FLE3}
    \end{split}    
\end{equation}

When differential operators are defined in time and space, for example, in a diffusive type equation~\cite{ccelik2012crank}, the discretization must be done carefully using backward and forward Euler-type techniques or the Crank-Nicolson method to correctly relate time and space. However, in our case the procedure is easier because all terms can be indexed exclusively by time and Eq.~\eqref{eq:FLE3} can be approximated by dividing the time interval $[0,t_n]$ into $n$ small steps of width $h$, such that
$x_n\equiv x(t_n)$ and $x_{n-1}\equiv x(t_{n-1})$,
\begin{align}
    \begin{split}
        x_n & \left[ \frac{1}{h} + \frac{\gamma}{M} \right] - \frac{x_{n-1}}{h} = v_0 + \frac{\mathcal{A}}{M}B(t_n) + \frac{\mathcal{A}_{H}}{M}B_{H}(t_n) \\
        & - \frac{\zeta}{M\Gamma(2H-1)} \sum_{i=0}^{n-1} \int_{ih}^{(i+1)h} (t_n-\tau)^{2H-2} x(\tau) d\tau.
        \label{eq:FLE_dis1}
    \end{split}
\end{align}

For the integral term in Eq.~\eqref{eq:FLE_dis1}, we can use the extended mean value theorem $\int_a^b f(x)g(x)dx\!=\!g(c)\int_a^bf(x)dx$ for $c\in[a,b]$, $f(x)$ continuous and $g(x)$ integrable and nonnegative on $[a,b]$ \cite{trench2003introduction}, to derive the approximate form 
\begin{multline}
    \int_{ih}^{(i+1)h} (t_n-\tau)^{2H-2} x(\tau) d\tau \\
     = \frac{x_i + x_{i+1}}{2}\int_{ih}^{(i+1)h}(t_n-\tau)^{2H-2}d\tau \\
     = \frac{x_i + x_{i+1}}{2(2H-1)}\left[(t_n -ih)^{2H-1} - (t_n-(i+1)h)^{2H-1}\right] \\
     \approx \frac{h^{2H-1}[x_i + x_{i+1}]}{2(2H-1)}\left[(n-i-1)^{2H-1} - (n-i-2)^{2H-1}\right],
\label{eq:FLE_dis2}
\end{multline}
where we truncate $t_n$ in the last line, to the $(n-1)th$ term, such that $t_n \approx (n-1)h$ for $n$ large enough. 

Finally, defining the index $j=n-i-1$, for $j\geq 1$ and replacing the approximate form of the integral~\eqref{eq:FLE_dis2}, we obtain the approximate recursive expression
\begin{align}
\begin{split}
    x_n & = \frac{Mh}{M+h\gamma}  \left[ \frac{1}{h}x_{n-1} + v_0 + \frac{\mathcal{A}}{M}B(t_n) + \frac{\mathcal{A}_{H}}{M}B_{H}(t_n)\right] \\
        & -\frac{\zeta h^{2H}}{M+h\gamma}\sum_{j=1}^{n-1} \frac{a_j \left[ x_{n-j} + x_{n-j-1} \right]}{2(2H-1)\Gamma(2H-1)},
\label{eq:FLE_dis3}
\end{split}
\end{align}
where $a_j = j^{2H-1} - (j-1)^{2H-1}$.

\section{\label{sec:fLe-sol}Analytical solutions of the fLe}

Analytical solutions of the fLe~\eqref{eq:fLe1} and expressions for the MSD of the system are obtained following standard mathematical techniques applied to Langevin-type equations~\cite{kwok2018langevin}. Taking the Laplace transform of Eq.~\eqref{eq:fLe1}, we obtain the solution for the position
\begin{equation}
    x(t) = x_0 + v_0 G(t) + \frac{1}{M} G(t)*f(t),
\label{eq:fLe_sol1}
\end{equation}
and the corresponding velocity
\begin{equation}
    v(t) = v_0 g(t) + \frac{1}{M} g(t)*f(t),
    \label{eq:fLe_sol2}
\end{equation}
where $x_0$ and $v_0$ are the initial position and velocity of the particle, respectively, and $*$ is the convolution operator. The functions $G(t)$ and $g(t) = dG(t)/dt$ in Eqs.~\eqref{eq:fLe_sol1} and \eqref{eq:fLe_sol2} are the relaxation functions of the system. $G(t)$ measures how the displacement of the particle forgets the initial velocity and $g(t)$ measures how the velocity forgets its initial value. This is easily observed by multiplying Eqs.~\eqref{eq:fLe_sol1} and~\eqref{eq:fLe_sol2} by $v_0$ and taking the expected value of the results, which leads to the following set of expressions:
\begin{align}
  \begin{split}
    G(t) & = \frac{\left< (x(t)-x_0)v_0 \right>}{\left< v^2_0 \right>}, \\
    g(t) & = \frac{\left< v_0v(t) \right>}{\left< v^2_0 \right>}.
    \label{eq:relaxation}
  \end{split}
\end{align}
These functions are defined by the inverse Laplace transform of
\begin{equation}
    \hat{G}(s) = \frac{1}{s^2 + \frac{\gamma}{M} s + \frac{\zeta}{M} s^{\alpha}}.
    \label{eq:G(s)}
\end{equation}

Using the Prabhakar Mittag-Leffler function \cite{garra2018prabhakar},
\begin{equation}
    E_{a,b}^r(z) = \frac{1}{\Gamma(r)}\sum_{k = 0}^{\infty}\frac{\Gamma(r+k)z^k}{k!\Gamma(ak+b)}
    \label{eq:MLP}
\end{equation}
together with the inverse Laplace transform obtained in Refs.~\cite{haubold2011mittag,saxena2004unified}, 
\begin{align}
\begin{split}
  & \mathscr{L}^{-1} \left\{ \frac{s^{\rho-1}}{s^{\mu} + a s^{\nu} + b}; t\right\} =\\ 
  & t^{\mu - \rho} \sum_{n=0}^{\infty} (-a)^n t^{(\mu-\nu)n}E_{\mu,\mu+(\mu-\nu)n-\rho+1}^{n+1} (-bt^{\mu}),  
\end{split}
\end{align}
for $\mathcal{R}(\mu)>0$, $\mathcal{R}(\nu)>0$, $\mathcal{R}(\rho)>0$ and $|as^{\nu}/(s^{\mu}+b)|<1$, we find
\begin{align}
    \begin{split}
        G(t) &= \mathscr{L}^{-1}\left\{ \hat{G}(s); t\right\} \\
             &= \sum^{\infty}_{n=0} \left(-\frac{\zeta}{M}\right)^n t^{(2-\alpha)n+1} E^{n+1}_{1, (2-\alpha)n +2}\left(-\frac{\gamma}{M} t\right).
        \label{eq:G(t)}
    \end{split}
\end{align}

\subsection{\label{sec:msd}Mean squared displacement (MSD)}
From Eq.~\eqref{eq:fLe_sol1}, we have
\begin{equation}
\begin{split}
    \langle x^2 (t) \rangle & =  \langle v_0^2 \rangle G^2(t) + I(t),
\end{split}
    \label{eq:MSD1}
\end{equation}
where we assume $x_0 = 0$ as the initial position. The function $I(t)$ represents the square of the convolution in Eq.~\eqref{eq:fLe_sol1}. Its general form is defined in terms of the correlation function $C(|\tau_1-\tau_2|) = \langle f(t-\tau_1)f(t-\tau_2)\rangle$ and the double integral over $d\tau_1$, $d\tau_2$. For simplicity in the calculations, we rewrite it by splitting the integrals into two areas separated at $\tau_1 = \tau_2$ and changing variables $\tau_1\leftrightarrow \tau_2$,
\begin{equation}
  \begin{split}
  I(t) & = \frac{1}{M^2} \int_0^t \int_0^t G(\tau_1) G(\tau_2) C(|\tau_1-\tau_2|) d\tau_2 d\tau_1\\   & = \frac{2}{M^2} \int_0^t G(\tau_1) \int_0^{\tau_1} G(\tau_2) C(\tau_1-\tau_2) d\tau_2 d\tau_1.
    \label{eq:I(t)}    
  \end{split}
\end{equation}

Under the specific conditions given by Eqs.~\eqref{eq:FDT} and \eqref{eq:FDT_weak}, the integral in $d\tau_2$ in Eq.~\eqref{eq:I(t)} can be simplified using the relationship between the time correlations of noise and the memory kernel of the system. To see this, we express the Laplace transform of the relaxation function \eqref{eq:G(s)} as
\begin{equation}
    \frac{1}{M}\hat{G}(s)\hat{\eta}(s) = \frac{1}{s} - s \hat{G}(s),
    \label{eq:etaG}
\end{equation}
where $\hat{\eta}(s) = \gamma + \zeta s^{\alpha - 1}$. Taking the inverse Laplace transform of Eq.~\eqref{eq:etaG}, we obtain
\begin{equation}
    \frac{1}{M} \int_0^tG(\tau)\eta(t-\tau) d\tau = 1 - g(t).
    \label{eq:etaG_conv}
\end{equation}

For the system described by linear and fractional friction terms, we use the fluctuation-dissipation relation given by Eq.~\eqref{eq:FDT} and Eq.~\eqref{eq:etaG_conv} to replace the integral in $d\tau_2$ in Eq.~\eqref{eq:I(t)} by $k_BT\left[1 - g(\tau_1)\right]$,
\begin{equation}
  \begin{split}
  I(t) = \frac{2 k_BT }{M}\left[\int_0^t G(\tau_1) d\tau_1- \frac{1}{2}G^2(t)\right],
    \label{eq:I(t)_eq}    
  \end{split}
\end{equation}
where $G(t=0)=0$ and $T=T_1=T_2$. Thus, assuming thermal initial conditions $\langle v_0^2\rangle = k_BT/M$ and replacing the  relaxation function \eqref{eq:G(t)}, we obtain
\begin{equation}
\begin{split}
    &\langle x^2  (t) \rangle = \frac{2 k_BT }{M}\int_0^t G(\tau_1) d\tau_1 \\
    &= \frac{2 k_BT }{M} \sum^{\infty}_{n=0} \left(-\frac{\zeta}{M} \right)^n t^{(2-\alpha)n+2} E^{n+1}_{1, (2-\alpha)n + 3}\left(-\frac{\gamma}{M} t\right).
    \label{eq:MSD_eq}
\end{split}
\end{equation}

The result in Eq.~\eqref{eq:MSD_eq} is a special case of the solution of the generalized Langevin equation for a free-particle driven by a mixture of $n$ power-law type noises, derived in Ref.~\cite{sandev2014langevin}, for the sub-diffusive regime $0<\alpha<1$. Furthermore, it agrees with the solutions obtained in Ref.~\cite{tateishi2012different} by means of the diffusive-like equation related to the generalized Langevin equation with white and colored noise.  The long-term behavior of the MSD in Eq.~\eqref{eq:MSD_eq} can be studied by means of the asymptotic expansion presented in Ref.~\cite{sandev2014langevin}, where it is shown that for long times $\langle x^2 \rangle \propto t^{\alpha}$.

\subsection{\label{sec:msd_neq}Zero linear friction cases}

In the limit $\gamma \to 0$, only orders $k=0$ in the Prabhakar Mittag-Leffler function $E^{n+1}_{1, (2-\alpha)n +2}\left(-\gamma t /M\right)$, given by Eq.~\eqref{eq:MLP}, survive for every $n = 0,1,...,\infty$. Thus, the relaxation function~\eqref{eq:G(t)} can be expressed as
\begin{align}
    \begin{split}
        \tilde{G}(t) & = \lim_{\gamma \to 0} G(t) \\
        & = \sum^{\infty}_{n=0} \left(-\frac{\zeta}{M}\right)^n \frac{t^{(2-\alpha)n+1}}{\Gamma\left[(2-\alpha)n + 2 \right]} \\
        & = t E_{2-\alpha,2}\left(-\frac{\zeta}{M} t^{2-\alpha}\right).
        \label{eq:G(t)_neq}
    \end{split}
\end{align}
Note that in the last line of Eq.~\eqref{eq:G(t)_neq}, we recover the two-parameter Mittag-Leffler function, which is defined as
\begin{equation}
    E_{a,b}(z) = \sum_{m = 0}^{\infty}\frac{z^m}{\Gamma(am+b)}.
    \label{eq:ML}
\end{equation}

The relaxation function~\eqref{eq:G(t)_neq} is also valid if we remove the white noise bath by setting $\eta_1=0$. In that case, the system is described by the conventional fLe with fractional friction and colored noise studied in Ref.~\cite{lutz2001fractional}, for which the second fluctuation-dissipation theorem is satisfied, such that Eqs.~\eqref{eq:FDT} and \eqref{eq:FDT_weak} are the same.

In the absence of linear friction, Eqs.~\eqref{eq:etaG} and \eqref{eq:etaG_conv} cannot be used directly to replace the whole integral in $d\tau_2$ in Eq.~\eqref{eq:I(t)}. In this case, it is easier to separate the correlations into white and colored components to analytically treat the MSD,
\begin{equation}
  \begin{split}
  I(t) & = \frac{2}{M^2} \int_0^t G(\tau_1) \int_0^{\tau_1} G(\tau_2) C_1(\tau_1-\tau_2) d\tau_2 d\tau_1 \\
  & + \frac{2}{M^2} \int_0^t G(\tau_1) \int_0^{\tau_1} G(\tau_2) C_2(\tau_1-\tau_2) d\tau_2 d\tau_1.
    \label{eq:I(t)_neq}    
  \end{split}
\end{equation}
The time correlation of the white noise \eqref{eq:corr_f1} can be replaced in the first term on the right-hand side of Eq.~\eqref{eq:I(t)_neq} to cancel out the integral in $d\tau_2$ using the Dirac delta, while the second term on the right-hand side can be simplified using the relationship between the memory kernel and the correlation time of the colored noise given by Eq.~\eqref{eq:FDT_weak}. Thus, replacing the integral in $d\tau_2$ over the colored noise correlation by $\eta_2k_BT_2\left[1 - g(\tau_1)\right]/\zeta$, we obtain
\begin{equation}
  \begin{split}
  \tilde{I}(t) & = \frac{2\eta_1 k_B \tilde{T}_1 t_\alpha^{1-\alpha}}{M^2} \int_0^t \tilde{G}^2(\tau_1) d\tau_1 \\
  & + \frac{2\eta_2 k_BT_2 }{\zeta M}\left[\int_0^t \tilde{G}(\tau_1) d\tau_1- \frac{1}{2}\tilde{G}^2(t)\right].
    \label{eq:I(t)_neq_2}    
  \end{split}
\end{equation}

For short time scales $t \ll (M/\zeta)^{1/2-\alpha}$, only small orders $n=0$ are relevant in the relaxation function \eqref{eq:G(t)_neq}, such that $\tilde{G}(t) \approx t/\Gamma(2)$. Thus, Eq.~\eqref{eq:I(t)_neq_2} behaves as
\begin{equation}
  \begin{split}
  \tilde{I}(t) & = \frac{2\eta_1 k_B \tilde{T}_1 t_\alpha^{1-\alpha}}{3M^2}t^3,
    \label{eq:I(t)_neq_st}    
  \end{split}
\end{equation}
where the second term on the right-hand side cancels out. Consequently, the MSD for short time scales is dominated by the ballistic behavior
\begin{equation}
\begin{split}
    \langle x^2  (t)\rangle_{st} & =  \langle v_0^2 \rangle \tilde{G}^2(t) + \tilde{I}(t) = \langle v_0^2 \rangle t^2 + \mathcal{O}(t^3).
\end{split}
    \label{eq:MSD_neq_st}
\end{equation}

We use the asymptotic relation for the two-parameter Mittag-Leffler function $E_{a,b}(-z) = 1/[z\Gamma(b-a)]$~\cite{mainardi2000mittag} to study the long-term behavior of Eq.~\eqref{eq:I(t)_neq_2}. After replacing $\tilde{G}(t)$ in Eq.~\eqref{eq:I(t)_neq_2}, we obtain
\begin{equation}
  \begin{split}
  \tilde{I}(t) & = \frac{2\eta_1 k_B \tilde{T}_1 t_\alpha^{1-\alpha}}{M^2} \int_0^t \left[\tau E_{2-\alpha,2}\left(-\frac{\zeta}{M} \tau^{2-\alpha}\right) \right]^2 d\tau \\
  & + \frac{2\eta_2 k_BT_2 }{\zeta M}\biggl\{t^2 E_{2-\alpha,3}\left(-\frac{\zeta}{M} t^{2-\alpha}\right) \\
  & - \frac{1}{2}\left[t E_{2-\alpha,2}\left(-\frac{\zeta}{M} t^{2-\alpha}\right) \right]^2\biggr\}.
    \label{eq:I(t)_neq_3}    
  \end{split}
\end{equation}
The long-term limit of the Mittag-Leffler function can be directly used for the terms outside the integral in Eq.~\eqref{eq:I(t)_neq_3}, which come from the contribution of colored noise. Replacing this asymptotic relation in the second and third terms of Eq.~\eqref{eq:I(t)_neq_3}, we obtain
\begin{equation}
  \begin{split}
  \frac{2\eta_2 k_BT_2}{\zeta M}&\biggl\{ t^2 E_{2-\alpha,3}\left(-\frac{\zeta}{M} t^{2-\alpha}\right) \\ & - \frac{1}{2} \left[t E_{2-\alpha,2}\left(-\frac{\zeta}{M} t^{2-\alpha}\right) \right]^2\biggr\}\\ 
   \approx \frac{2 \eta_2 k_BT_2 }{\zeta^2} & \biggl[\frac{1}{\Gamma(1+\alpha)} t^{\alpha} - \frac{M}{2\zeta\Gamma(\alpha)^2} t^{2\alpha-2}\biggr] .
    \label{eq:I(t)_neq_st_c}    
  \end{split}
\end{equation}
However, the integral that comes from the contribution of white noise must be evaluated carefully. For the long time scale $t>t_l$, the first term of Eq.~\eqref{eq:I(t)_neq_3} reads
\begin{equation}
  \begin{split}
  &\frac{2\eta_1 k_B \tilde{T}_1 t_\alpha^{1-\alpha}}{M^2} \int_{t_l}^t \left[\tau_1 E_{2-\alpha,2}\left(-\frac{\zeta}{M} \tau_1^{2-\alpha}\right) \right]^2 d\tau_1 \\ 
  & = \frac{2\eta_1 k_B \tilde{T}_1 t_\alpha^{1-\alpha}}{\zeta^2\Gamma(\alpha)^2}
  \begin{cases}
      \frac{1}{2\alpha-1}(t^{2\alpha -1}-t_l^{2\alpha -1})  & \text{if } \alpha \neq 0.5, \\
       \log(t/t_l) & \text{if } \alpha = 0.5.
    \end{cases}
    \label{eq:I(t)_neq_st_w}
  \end{split}
\end{equation}

Again, if we assume thermal initial conditions over the bath in equilibrium $\langle v_0^2 \rangle = k_BT_2/M$, the dependence on $\tilde{G}^2(t)$ cancels out for the MSD in Eq.~\eqref{eq:MSD1}. Replacing Eqs.~\eqref{eq:I(t)_neq_st_c} and \eqref{eq:I(t)_neq_st_w} in \eqref{eq:MSD1}, we see that the dominant order in the long-term behavior is 
\begin{equation}
\begin{split}
    \langle x^2(t) \rangle_{lt} =  \frac{2\eta_2 k_BT_2 }{\zeta^2\Gamma(1+\alpha)}
    \begin{cases}
       t^{\alpha} + \mathcal{O}(t^{2\alpha-1})  & \text{if } \alpha \neq 0.5, \\
       \sqrt{t} + \mathcal{O}(\log(t)) & \text{if } \alpha = 0.5.
    \end{cases}
    \label{eq:MSD_neq_lt}
\end{split}
\end{equation}
The dominant order for the long-term MSD of the zero-linear friction case found here is $t^\alpha$, as for the in-equilibrium MSD given by Eq.~\eqref{eq:MSD_eq} and for the fLe with fractional friction and colored noise studied in Ref.~\cite{lutz2001fractional}. In all these cases, the dominant contribution comes from the colored-noise thermal bath. On the other hand, when bath $2$ is removed, the MSD is dominated in the long term by $t^{2\alpha -1}$, as in Ref.~\cite{verstraten2021time}.

\section{\label{sec:results}Results}

\begin{figure*}[ht]
    \centering
    \includegraphics[scale=0.47]{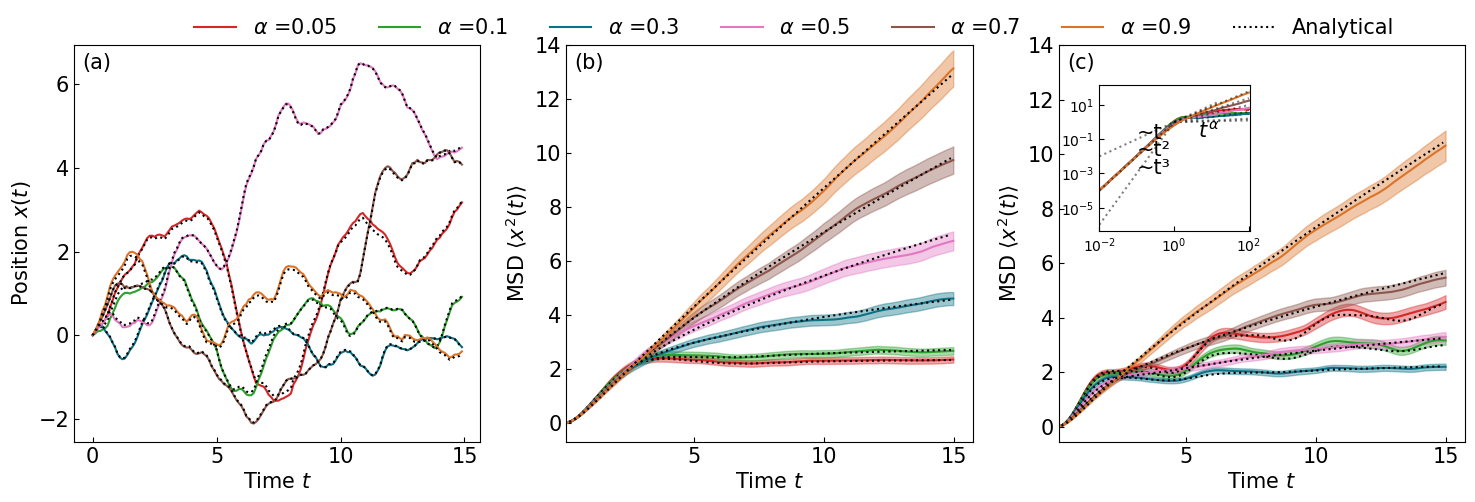}
    \caption{Analytical and numerical results for the position and the MSD. Solid colored lines were obtained from the numerical implementation of the discrete recursive expression in Eq.~\eqref{eq:FLE_dis3} for orders $\alpha = \{0.05, 0.1, 0.3, 0.5, 0.7, 0.9\}$, and shaded areas are the $98\%$ normal confidence intervals. Analytical solutions are plotted in black dotted lines and asymptotic limits in gray dotted lines. (a) Solutions for the position of the system with both friction terms and both noise baths. (b) MSD of the system with both friction terms and both noise baths. (c) MSD of the zero linear friction case ($\gamma \to 0$). The short and long-term asymptotic behavior are shown in the inset.}
    \label{fig:FE}
\end{figure*}

We performed numerical calculations for the position of the system using the recursive discrete expression given by Eq.~\eqref{eq:FLE_dis3}. The fBm trajectories were generated using the algorithms discussed in Appendix~\ref{AppNoise}, setting the normalization constant $\Lambda=1$. The numerical MSD was obtained by explicitly computing $(x - x_0)^2$, using $x_0=0$. In order to deal with the propagation of error that affected small orders of $\alpha$, we used two different time-step sizes, h=$0.005$ for $\alpha \leq 0.1$, and a higher one $h=0.01$ for orders $\alpha>0.1$ to speed up the algorithm. For $h=0.01$, we averaged over 16000 solutions and for $h=0.005$ over 4000 solutions. Normal confidence intervals of $98\%$ were calculated for all the cases. We normalized all physical constants to 1, such that $M=\eta_{1,2}=k_BT_{1,2}=1$, for the system with fractional- and linear-order friction terms, and $M=\eta_{1,2}=k_B\tilde{T}_1=k_BT_2=1$, for the zero linear friction case. This last case implies that the local temperature in the thermal gradient satisfies $\tilde{T}_1/T_1 = \sin(\pi\alpha/2)$. By fine-tuning the temperatures, we homogenized the thermal initial condition $\langle v_0^2\rangle=k_B{T_1}/M = k_B{T_2}/M$ and $\langle v_0^2\rangle=k_B{\tilde{T}_1}/M = k_B{T_2}/M$, for each case, respectively. Additionally, it allowed us to fix the initial velocity $v_0 = \sqrt{\langle v_0^2\rangle}=1$, to avoid sampling this value from an arbitrary distribution.

In Figs.~\ref{fig:FE}(a) and \ref{fig:FE}(b), we present the analytical (dotted lines) and the numerical (solid lines) results with confidence intervals (shaded areas), for the position and the MSD of the system described by both friction terms and both noise terms, while the MSD of the case without linear friction is presented in Fig.~\ref{fig:FE}(c). The curves for the analytical expressions of the position were generated by evaluating Eq.~\eqref{eq:fLe_sol1}, with the relaxation function given by Eq.~\eqref{eq:G(t)} and the analytical curves of the MSD were calculated by evaluating Eqs.~\eqref{eq:MSD_eq} and \eqref{eq:I(t)_neq_st_c}, respectively. In particular, to evaluate the contribution of white noise in the MSD of the system without linear friction, we numerically integrated the square of the relaxation function given by Eq.\eqref{eq:G(t)_neq}. The short and long-term expressions were evaluated using Eqs.~\eqref{eq:MSD_neq_st} and \eqref{eq:MSD_neq_lt}, respectively. All the cases evaluated exhibit a very good agreement between analytical and numerical solutions.

\subsection{\label{sec:msd_tc}Time crystal phases}

\begin{figure*}[ht]
    \centering
    \includegraphics[scale=0.47]{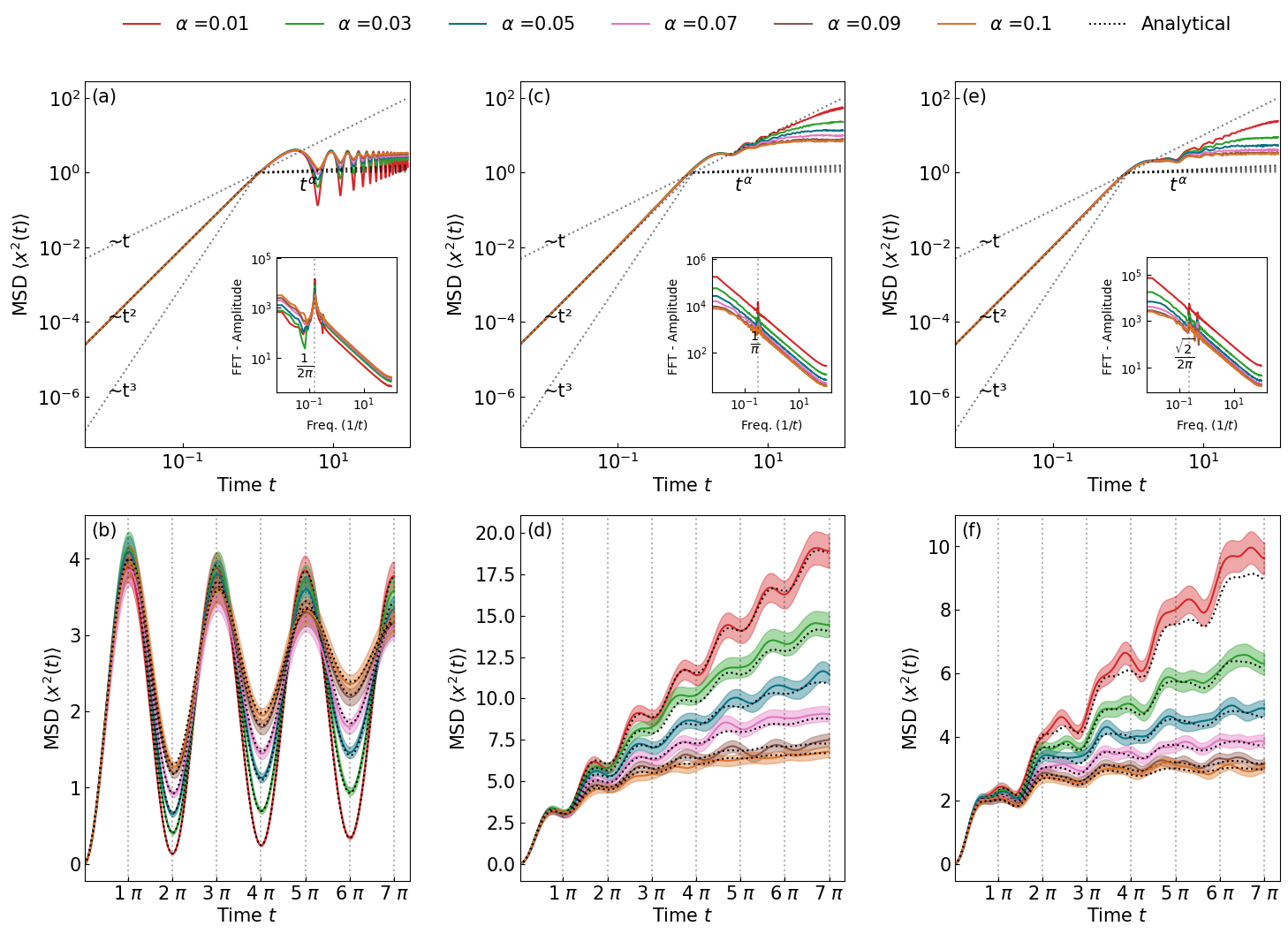}
    \caption{Periodic time-ordered phases emerge in the MSD from the system with zero linear friction in the regime $\alpha \lesssim 0.1$. Solid colored lines were obtained from the numerical implementation of the discrete recursive expression in Eq.~\eqref{eq:FLE_dis3}, and shaded areas are the $98\%$ normal confidence intervals. Analytical solutions are plotted in black dotted lines and asymptotic limits in gray dotted lines. In (a), (c), and (e), the MSD is plotted versus time in a log-log scale to present the short- and long-term regimes, whereas in (b), (d), and (f), a linear scale is used to better reveal the oscillations. The time crystal phase driven by the action of the colored noise with dominant periodicity of approximately $2\pi$ can be observed in (a) and (b). The time glass phase driven by the white noise with dominant periodicity proportional to approximately $\pi$ can be observed in (c) and (d). In (e) and (f), the mixed phase created by the superposition of the white and colored noises is shown. The dominant periodicity of the mixed phase is proportional to approximately $\sqrt{2}/2\pi$. The insets of figures (a), (c), and (e) show the spectral analysis of the MSD for all the phases performed using FFT.}
    \label{fig:TC}
\end{figure*}

The absence of linear friction ($\gamma \to 0$) in the system triggers the emergence of periodic time-ordered phases in the MSD for $\alpha \lesssim 0.1$. In Fig.~\ref{fig:FE}(c), we observe oscillations that show up for $\alpha = 0.05$ and $\alpha = 0.1$. Although the long-term behavior is still dominated by $t^{\alpha}$, in this regime the MSD reaches higher values as $\alpha$ decreases. A detailed analysis of the different periodic-time ordered phases that emerge in the small-orders regime $\alpha \lesssim 0.1$ is presented in Fig.~\ref{fig:TC}. This is the main result of our work.

When the white noise bath is neglected ($\eta_1 = 0$), the system exhibits the signature of a time crystal. A ground state emerges from thermal equilibrium, satisfying the fluctuation-dissipation theorem, and the MSD manifests an emergent periodic ordering. In Fig.~\ref{fig:TC}(a), we observe that the short-term regime of the time crystal phase displays ballistic behavior, while, in the long-term, damped oscillations occur. The characteristic frequency of oscillations for different fractional orders converges to approximately $2\pi$, as it is shown in Fig.~\ref{fig:TC}(b). 

In contrast to the previous case, when the colored noise bath is neglected ($\eta_2 = 0$), the system exhibits the characteristics of the time glass revealed in Ref.~\cite{verstraten2021time}. The fluctuation-dissipation theorem is broken and the periodicity of the emergent time-ordered phase converges to approximately $\pi$. In Fig.~\ref{fig:TC}(c), we observe that the ballistic behavior holds in the short-term regime. However, in the long term, the oscillations display periodic plateaus until the system saturates, due to the negative exponent in the asymptotic limit $t^{2\alpha-2}$. A detailed view of the periodic plateaus up to $5\pi$ is presented in Fig.~\ref{fig:TC}(d). In particular, we observe that the shape of the plateaus is better defined when the initial velocity of the system is smaller. This wanes the additional oscillatory effect of the square of the relaxation function in Eq.~\ref{eq:MSD1}, in absence of colored noise.

The action of both baths generates a mixed time-ordered phase, where the dominant periodicity becomes $\sqrt{2}\pi$. In the short-term, the universal ballistic behavior, also exhibited by the other phases, is preserved. Furthermore, the sub-diffusive long-term limit, proportional to $t^{\alpha}$, is recovered by the system, due to the action of the colored noise. Both regimes are shown in Fig.~\ref{fig:TC}(e). Interestingly, the features of both, time glass and time crystal can be observed in the oscillatory regime. Small plateaus appear at the beginning of each period, followed by damped oscillations. The detail of this phase is shown in Fig.~\ref{fig:TC}(f).

For all the three phases studied, we numerically check the value of the periodicity by calculating the dominant frequencies from the spectral analysis of the MSD using the Fast Fourier Transform (FFT) technique. The results are presented in the insets of Fig.~\ref{fig:TC}. To better understand the emergent periodicity of the system, we evaluate the limit $\alpha \to 0$ in the relaxation function given by Eq.~\eqref{eq:G(t)_neq}. Using the special case of the Mittag-Leffler function $E_{2,2} (z) = \sinh(\sqrt{z})/\sqrt{z}$ \cite{haubold2011mittag}, we obtain
\begin{equation}
  \begin{split}
    \lim_{\alpha \to 0} \tilde{G}(t) & = \sqrt{-\frac{M}{\zeta}}\sinh \left(\sqrt{-\frac{\zeta}{M}} t\right) \\
    & = \sqrt{\frac{M}{\zeta}}\sin \left(- \sqrt{\frac{\zeta}{M}} t\right),
  \label{eq:G(t)_neq_0}
  \end{split}
\end{equation}
where we used $\sinh(t)=-i\sin(it)$.

Inserting this relaxation function in Eq.~\eqref{eq:MSD1}, we find
\begin{equation}
  \begin{split}
  \langle x^2(t) \rangle & = \frac{\eta_1 k_B \tilde{T}_1 t_\alpha^{1-\alpha}}{M\zeta} \left[ t - \frac{1}{2}\sqrt{\frac{M}{\zeta}}\sin\left(2\sqrt{\frac{\zeta}{M}} t \right)\right] \\
  & + \frac{2 \eta_2 k_BT_2 }{\zeta^2}\left[\cos\left(\sqrt{\frac{\zeta}{M}} t \right) -1 \right],
    \label{eq:MSD_neq_0}    
  \end{split}
\end{equation}
where we assume the thermal initial condition $\langle v_0^2 \rangle = k_BT_2/M$. The first term of the MSD comes from the contribution of the white noise bath; the periodicity of this term is $\pi\sqrt{M/\zeta}$. On the other hand, the second term is driven by the colored noise and its periodicity is $2\pi\sqrt{M/\zeta}$. When $\alpha \neq 0$, the periodicity of these terms is $P_1 = \pi(M/\zeta)^{1/(2-\alpha)}$ and $P_2 = 2P_1$. Thus, the general periodicity of the time glass phase is given by $P_1$, while the dominant periodicity of the time crystal and the mixed phase is given by $P_2$. In particular, we observe that the superposition of the damping effects of both baths $\zeta = \eta_1 + \eta_2$, given by Eq.~\eqref{eq:eta_neq} deviates the periodicity of the mixed phase from the time crystal phase.

\section{\label{sec:derivation} Microscopic derivation of fLe}

In the previous sections, we discussed the main macroscopic features of the fLe \eqref{eq:fLe1}, pictured as a particle coupled to two different thermal baths causing the noises and the frictions in the system. In this section, we investigate the microscopic model to formally derive the fLe and show how all the studied combinations of parameters (and thus all physical regimes) emerge. 

In the Caldeira-Leggett model, a single particle is coupled to an environment, such that all degrees of freedom are assumed to be in a local minimum associated to thermal equilibrium . Expanding around these to second order, we can describe this general environment in terms of a set of harmonic oscillators. These harmonic oscillators can then be integrated out to yield an effective semi-classical description of the system of interest. Following Ref.~\cite{caldeira2014introduction}, we start with a Lagrangian of the form
\begin{equation}
    L = L_S + \sum_{i \in \{ 1,2 \} } \Big( L_{B_i} + L_{I_i} + L_{CT_i} \Big),
\end{equation}
where $L_S$ is the Lagrangian of the system, $L_{B_i}$ is the Lagrangian of the bath, $L_{I_i}$ is the interaction between system and bath, and $L_{CT_i}$ is a counter term, with the subindex $i={1,2}$ representing each of the two baths. In particular, we have
\begin{align}
    L_S &= \frac{1}{2} M \dot{x}^2 -V(x),\\
    L_{B_i} &= \sum_j \frac{1}{2}m_{i,j} \dot{q}_{i,j}^2 -\frac{1}{2} m_{i,j} \omega_{i,j}^2 q_{i,j}^2,\\
    L_{I_i} &= \sum_j C_{i,j} q_{i,j} x, \\
    L_{CT_i} &= -\sum_{j} \frac{C_{i,j}^2}{2 m_{i,j} \omega_{i,j}^2} x^2 ,
\end{align}
where $x$ is the position of a particle with mass $M$ in a potential $V(x)$, $m_{i,j}$ is the mass of the harmonic oscillators at position $q_{i,j}$ and frequency $\omega_{i,j}$, and $C_{i,j}$ is the coupling strength between the particle and the $\{i,j\}^{th}$ harmonic oscillator. The counter term is there to compensate the potential shift in $V(x)$ due to the presence of the bath, making sure that the effective potential remains equal to $V(x)$.

We can take the classical Euler-Lagrange equations of motion, which result in
\begin{align}
    M \ddot{x} &=-V'(x)+\sum_{i,j} C_{i,j} q_{i,j} -\frac{C_{i,j}^2}{m_{i,j} \omega_{i,j}^2} x, \label{eq:eomsys} \\
    m_{i,j} \ddot{q}_{i,j} &= -m_{i,j} \omega_{i,j}^2 q_{i,j} + C_{i,j} x. \label{eq:eombath}
\end{align}
We solve the latter equation for $q_{i,j}(t)$ as a function of $x(t)$ by Laplace transforming, which becomes
\begin{align}
        q_{i,j}(s) &= \frac{\dot{q}_{i,j}(0)}{s^2+\omega_{i,j}^2} + \frac{s q_{i,j}(0)}{s^2+\omega_{i,j}^2} \nonumber \\
        &+\frac{C_{i,j} x(s)}{m_{i,j}} \left( \frac{1}{\omega_{i,j}^2}-\frac{1}{\omega_{i,j}^2}\frac{s^2}{s^2+\omega_{i,j}^2} \right),
        \intertext{and consequently}
        q_{i,j}(t)&= \frac{\dot{q}_{i,j}(0)}{\omega_{i,j}}\sin(\omega_{i,j} t) +q_{i,j}(0) \cos(\omega_{i,j} t) \nonumber\\
        +& \frac{C_{i,j} x(t)}{m_{i,j} \omega_{i,j}^2} -\frac{d}{dt} \left[ \frac{C_{i,j}}{m_{i,j}\omega_{i,j}^2} x(t) *\cos(\omega_{i,j} t) \right].
\end{align}
Next, we can insert these solutions for the bath into Eq.~\eqref{eq:eomsys}, yielding
\begin{align}
    M \ddot{x} 
    &= -V'(x) -\sum_{i,j} \frac{d}{dt} \left[ \frac{C_{i,j}^2}{m_{i,j}\omega_{i,j}^2} x(t) *\cos(\omega_{i,j} t) \right]  \nonumber\\
        &+\sum_{i,j}C_{i,j} \Bigg\{\frac{\dot{q}_{i,j}(0)}{\omega_{i,j}}\sin(\omega_{i,j} t) +q_{i,j}(0) \cos(\omega_{i,j} t)  \Bigg\}. \label{eq:bigeom}
\end{align}
Here, we see two new terms arising due to the presence of the bath: the first term is what will determine the dissipation, whereas the second term will cause fluctuations. In particular, we see that when the thermal equilibrium of the system is preserved, these terms lead to the emergence of a fluctuation-dissipation relation of the type \eqref{eq:FDT}.

In order to further simplify, we evoke the spectral function of the bath, which is given by the imaginary part of the retarded dynamical susceptibility $J_i(\omega)=\text{Im } \chi_{B_i}(\omega)$ of the two baths. Taking the Fourier transform of Eq.~\eqref{eq:eombath}, we see that $q_{i,j}(\omega)=C_{i,j} x(\omega)/m_{i,j}(\omega_{i,j}^2-\omega^2)$, which means that 
\begin{align}
    \chi_{B_i}(\omega) &= \sum_{j} C_{i,j} \frac{\delta q_{i,j}(\omega)}{\delta x(\omega)}= \sum_{j} \frac{C_{i,j}^2}{m_{i,j}(\omega_{i,j}^2-\omega^2)}\nonumber\\
    = \sum_{j}& \left( \frac{C_{i,j}^2}{2m_{i,j}\omega_{i,j}(\omega+\omega_{i,j})} - \frac{C_{i,j}^2}{2m_{i,j}\omega_{i,j}(\omega-\omega_{i,j})} \right).
\end{align}
Next, we need to add an infinitesimal shift from the real axis, which we do by shifting $\omega \pm \omega_{i,j}\to \omega \pm \omega_{i,j} + i\varepsilon$, with $\varepsilon\to 0$. With this, we can use the identity $\text{Im } 1/[(\omega\pm\omega_{i,j})+i\varepsilon] =-\pi \delta(\omega\pm \omega_{i,j})$ in order to compute the spectral functions as
\begin{align}
    J_i(\omega)&= \text{Im} \chi_{B_i}(\omega)= \frac{\pi}{2} \sum_{j} \frac{C_{i,j}^2}{m_{i,j}\omega_{i,j}} \delta(\omega-\omega_{i,j}), \label{eq:spectral}
\end{align}
where we assume that all the bath frequencies $\omega_{i,j}$ are positive. The bath spectral functions are the crucial connection between the microscopic model and the macroscopic physics, as it turns out that the only relevant detail of the microscopic bath is the strength at which the different frequencies couple to the system. This can be achieved in two different ways: Firstly, there is the distribution of the frequency modes in the bath and, secondly, there is the strength at which each mode couples to the system. Typically, a power-law $J_i(\omega) \propto \eta_i\omega^{\alpha_i}$, with $\alpha_i>0$, is assumed for the general shape of the spectral function~\footnote{The spectral function can be much more general. As such, one could also assume only a single spectral function $J(\omega)=\eta_1 \omega^{\alpha_1} +\eta_2 \omega^{\alpha_2}$. However, this would mean that $T_1=T_2$, which fixes the relative noise strength, whereas having two separate baths allows for different temperatures and thus a larger parameter space.}.
Thus, to find the macroscopic description, we want to write Eq.~\eqref{eq:bigeom} in terms of the two spectral functions. For the friction term, we find that
\begin{align}
   &F_{fr,i}(x) := \sum_{j} \frac{d}{dt} \left[ \frac{C_{i,j}^2}{m_{i,j}\omega_{i,j}^2} x(t) *\cos(\omega_{i,j} t) \right] \nonumber\\
    &=  \frac{d}{dt} \left[ \frac{2}{\pi} \int_0^t d\tau \int_0^\infty d\omega \frac{J_i(\omega)}{\omega} \cos[\omega(t-\tau)]x(\tau)\right],\label{eq:ffr}
\end{align}
where we converted the sum over $j$ to an integral over delta functions and expanded the convolution. The fluctuations can be captured by a noise term
\begin{align}
    f_i(t)&:= \sum_j C_{i,j} \Big[\frac{\dot{q}_{i,j}(0)}{\omega_{i,j}}\sin(\omega_{i,j} t) +q_{i,j}(0) \cos(\omega_{i,j} t)  \Big],\label{eq:fi}
\end{align}
where we treat $\dot{q}_{i,j}(0)$ and $q_{i,j}(0)$ as stochastic variables. 
Combining all these into Eq.~\eqref{eq:bigeom}, we conclude that
\begin{align}
    M \ddot{x}+V'(x) +F_{fr,1}(x) +F_{fr,2}(x) &=  f_1(t) +f_2(t). \label{eq:generalfle}
\end{align}

With our linear system-bath coupling, if we assume an Ohmic bath, we will find a linear spectral function, while a non-Ohmic bath will give rise to a non-linear spectral function~\cite{caldeira2014introduction}. However, recent work has shown that this non-linear spectral function can also be achieved with Ohmic baths if the coupling becomes fractional~\cite{vertessen2023dissipative}. 
Therefore, although the baths could be any power-law, we will assume bath 1 to be Ohmic $J_1(\omega)=\eta_1 \omega$, while keeping bath 2 more general, with a power-law
\begin{align}
    J_1(\omega)&=\eta_1 \omega,\label{eq:j1}\\
    J_2(\omega)&=\eta_2 \sin(\pi \alpha/2) \omega^\alpha,\label{eq:ja}
\end{align}
where we assume that $0<\alpha<1$. Inserting Eq.~\eqref{eq:j1} into Eq.~\eqref{eq:ffr} and noticing the delta function in the $\omega$ integral, we obtain the friction of bath $1$,
\begin{align}
    F_{fr,1}(x) &= 2\eta_1 \frac{d}{dt} \int_0^t d\tau \delta(t-\tau) x(\tau) =\eta_1 \dot{x}(t),\label{eq:ffr1}
\end{align}
where the factor of 2 disappeared due to the delta function being evaluated on the integral boundary. Similarly, for the friction  of bath $2$, we insert Eq.~\eqref{eq:ja} into Eq.~\eqref{eq:ffr}, and make a change of variables $\omega\to \omega/(t-t')$ in order to decouple the integrals,
\begin{align}
    &F_{fr,2} (x) = \frac{2\eta_2}{\pi} \sin\left(\frac{\pi \alpha}{2}\right)  
 \int_0^\infty d\omega \omega^{\alpha-1}\cos\omega \nonumber\\ 
 &\times\frac{d}{dt}\int_0^td\tau (t-\tau)^{-\alpha} x(\tau) \nonumber\\
 &=
 \frac{2\eta_2}{\pi} \sin\left(\frac{\pi \alpha}{2}\right) \cos\left(\frac{\pi \alpha}{2} \right) \Gamma(\alpha) \Gamma(1-\alpha) \prescript{RL}{0}{D}^{\alpha}_t x(t).
\end{align}

We can use Eq.~\eqref{eq:FC-C_RL_0} to rewrite the Riemann-Liouville fractional derivative as a Caputo operator, where we assume that $x(0)=0$. We can also use Euler's reflection formula and some trigonometric identities to simplify the prefactors and obtain
\begin{align}
    F_{fr,2} (x) &= \eta_2 \prescript{C}{0}{D}^{\alpha}_t x(t). \label{eq:ffralpha}
\end{align}

\subsection{In equilibrium noise}
The noise term in Eq.~\eqref{eq:fi} contains two independent stochastic variables, in the ensemble sense. In a classical equilibrium, these are determined by the equipartition theorem, which tells us that each average energetic degree of freedom, such as kinetic and potential energy, is related to the temperature. From this, we can derive the expectation values
\begin{align}
    \langle q_{i,j}(0)\rangle=\langle \dot{q}_{i,j}(0)\rangle&=\langle q_{i,j}(0) \dot{q}_{i,j}(0)\rangle=0,\label{eq:equi1}\\
 \langle \dot{q}_{i,j}(0) \dot{q}_{i',j'}(0)\rangle&= \frac{k_B T_i}{m_{i,j}} \delta_{i,i'}\delta_{j,j'},\label{eq:equi2}\\
 \langle q_{i,j}(0) q_{i',j'}(0)\rangle&= \frac{k_B T_i}{m_{i,j}\omega_{i,j}^2} \delta_{i,i'}\delta_{j,j'},\label{eq:equi3}
\end{align}
where $k_B$ is the Boltzmann constant and $T_i$ is the temperature of bath 1 or 2. Using these relations, we can determine the correlations and expectations of $f_i(t)$. From Eq.~\eqref{eq:equi1}, we can directly see that $\langle f_1(t) \rangle = \langle f_2(t) \rangle =0$. For the correlation, we find
\begin{align}
    &\langle f_i(t)f_i(t') \rangle =\sum_{j,j'} C_{i,j}C_{i,j'}\times \nonumber\\
    &\Bigg\langle \Big[\frac{\dot{q}_{i,j}(0)}{\omega_{i,j}}\sin(\omega_{i,j} t) +q_{i,j}(0) \cos(\omega_{i,j} t)  \Big]\nonumber\\
    &\Big[\frac{\dot{q}_{i,j'}(0)}{\omega_{i,j'}}\sin(\omega_{i,j'} t') +q_{i,j'}(0) \cos(\omega_{i,j'} t')  \Big] \Bigg\rangle \nonumber\\
    &= \sum_{j,j'} C_{i,j}C_{i,j'}  \Big[\frac{\langle \dot{q}_{i,j}(0) \dot{q}_{i,j'}(0)\rangle}{\omega_{i,j}\omega_{i,j'}}\sin(\omega_{i,j} t)\sin(\omega_{i,j'} t') \nonumber\\ 
    &+\langle q_{i,j}(0)q_{i,j'}(0)\rangle \cos(\omega_{i,j} t)\cos(\omega_{i,j'} t')  \Big],
\end{align}
where we already used that there is no correlation between $q$ and $\dot{q}$. We can now insert Eqs. \eqref{eq:equi2} and \eqref{eq:equi3} to get
\begin{align}
   &\langle f_i(t)f_i(t') \rangle = \sum_{j}  \frac{k_B T_i C_{i,j}^2}{m_{i,j}\omega_{i,j}^2} \times \nonumber\\ 
    &\Big[\sin(\omega_{i,j} t)\sin(\omega_{i,j} t') + \cos(\omega_{i,j} t)\cos(\omega_{i,j} t')  \Big]\\
    &= \sum_{j}  \frac{k_B T_i C_{i,j}^2}{m_{i,j}\omega_{i,j}^2} \cos[\omega_{i,j}(t-t')].
\end{align}

Finally, we can identify the spectral function, Eq.~\eqref{eq:spectral}, as
\begin{align}
    &\langle f_i(t)f_i(t') \rangle = \nonumber\\
    &k_B T_i\int_0^\infty d\omega \sum_j  \frac{ C_{i,j}^2}{m_{i,j}\omega_{i,j}}  \delta(\omega-\omega_{i,j}) \cos[\omega(t-t')]/\omega\nonumber\\
    &= \frac{2}{\pi} k_B T_i\int_0^\infty d\omega \frac{J_i(\omega)}{\omega} \cos[\omega(t-t')].\label{eq:equicorr}
\end{align}
For the Ohmic bath, we find that
\begin{align}
    \langle f_1(t)f_1(t') \rangle &=\frac{2}{\pi} k_B T_1 \eta_1\int_0^\infty d\omega  \cos[\omega(t-t')] \nonumber\\
    &= 2 \eta_1 k_B T_1 \delta(t-t'),
\end{align}
which is the white noise correlation associated to regular Brownian motion \cite{caldeira1983path}. For the non-Ohmic bath, we obtain
\begin{align}
    \langle &f_2(t) f_2(t') \rangle = \frac{2}{\pi} \eta_2 \sin\left(\frac{\pi \alpha}{2}\right) k_B T_2 \nonumber\\ 
    &\times\int_0^\infty d\omega \omega^{\alpha-1} \cos[\omega(t-t')] \nonumber\\
    &= \frac{2}{\pi} \eta_2 \sin\left(\frac{\pi \alpha}{2}\right) k_B T_2 |t-t'|^{-\alpha} \int_0^\infty d\omega \omega^{\alpha-1} \cos[\omega] \nonumber\\
    &= \frac{2}{\pi} \eta_2 \sin\left(\frac{\pi \alpha}{2}\right) k_B T_2 |t-t'|^{-\alpha}\cos\left(\frac{\pi \alpha}{2}\right) \Gamma(\alpha)\nonumber\\
    &= \eta_2 k_B T_2 \frac{|t-t'|^{-\alpha}}{\Gamma(1-\alpha)}. 
\end{align}
Combining all results together and setting $V(x)=0$, we conclude that the final equation of motion for the particle is given by
\begin{align}
      M \ddot{x}(t) +\eta_1 \dot{x}(t) &+\eta_2\prescript{C}{0}{D}^{\alpha}_t x(t)  =  f_1(t) +f_2(t),\\
      \langle f_1(t) f_1(t')\rangle &=  2 \eta_1 k_B T_1 \delta(t-t'),\\
      \langle f_2(t) f_2(t') \rangle &= \eta_2 k_B T_2 \frac{|t-t'|^{-\alpha}}{\Gamma(1-\alpha)}, \\
      \langle f_1(t) f_2(t') \rangle &= 0.
\end{align}

\subsection{Out of equilibrium noise}
The previous derivation of the fLe describes a system in thermal equilibrium, where every bath generates fluctuation-dissipation pairs: white noise-linear friction, colored noise-fractional friction. Thermal equilibrium holds if we neglect any of the baths, which yields to the conventional Langevin equation describing Brownian Motion and linear diffusion $\langle x^2(t) \rangle \propto t$; and the fractional Langevin equation with colored noise exhibiting sub-diffusion $\langle x^2(t) \rangle \propto t^\alpha$. However, in the case where we want to retain white noise but remove the linear friction, we need the white noise to come from a bath which has the same order as the second bath, related to $\alpha$. 

With that idea in mind, we assume that bath 2 remains unchanged, but bath 1 will now have the same $\alpha$-order spectral function: $J_1= \eta_1 \sin(\pi \alpha/2) \omega^\alpha $. In this way, the friction term will simply add up with the other fractional friction term. However, in order to retain white noise, we need to break equipartition. We do this in accordance with Ref.~\cite{verstraten2021time}, where a local temperature is introduced for each harmonic oscillator to apply a temperature gradient across the bath.  Formally, this is done by replacing 
\begin{equation}
    T_1\to T_{1} [t_\alpha \omega_{1,j}]^{1-\alpha},
\end{equation}
where $t_\alpha=(M/\eta_1)^{1/(2-\alpha)}$ and, importantly, the extra $j$-dependent factor is dimensionless. 
Analyzing the derivation in the equilibrium case, we see that, up to the new factor, nothing in the calculations changes up to Eq.~\eqref{eq:equicorr}, where in the out-of-equilibrium case have
\begin{align}
    \langle f_1(t)f_1(t') \rangle &= \frac{2 k_B T_1 t_\alpha^{1-\alpha}}{\pi } \int_0^\infty d\omega \frac{J_1(\omega)}{\omega^{\alpha}} \cos[\omega(t-t')].
\end{align}
Inserting the spectral function, we immediately see that the $\omega$ power drops out and we find a delta function in the cosine integral, yielding
\begin{align}
    \langle f_1(t)f_1(t') \rangle &= 2 \eta_1 k_B \tilde{T}_1  t_\alpha^{1-\alpha} \delta(t-t'),
\end{align}
where we define $\Tilde{T}_1= T_1 \sin\left(\pi\alpha/2\right)$ as the effective temperature of the bath.

\section{Conclusions}
\label{sec:con}

In this work, we solved analytically and numerically the fLe~\eqref{eq:fLe1} of a system coupled to two baths causing different noise terms and frictions. When the system is in equilibrium, each bath leads to a fluctuation-dissipation pair. In that scenario, we assume that bath $1$ induces linear friction and white noise. and bath $2$ causes fractional friction and colored noise. The analytical results for the position [Eq.~\eqref{eq:fLe_sol1}] and the MSD [Eq.~\eqref{eq:MSD_eq}] were expressed using the Prabhakar or three-parameter Mittag-Leffler function. For thermal initial conditions, the MSD is expressed by integrating the relaxation functions given by the Prabhakar-Mittag-Leffler. 

Numerical solutions of the system were obtained by modeling the thermal baths using fractional Brownian noise. In particular, we used the derivative of oBm to model white noise and the derivative of fBm for colored noise. This allowed us to integrate in time the fLe, in order to obtain recursive discrete expression that directly depend on the oBm and fBm trajectories. This approach constitutes a suitable method to easily and quickly solve Langevin-type equations. Its convergence has been discussed in Refs.~\cite{neuenkirch2007exact, hu2016rate}. However, for small fractional orders, the error propagation affects the results and demands decreasing the size of the step in the discretization, increasing the computational time for calculating the MSD.

The MSD exhibits a dominant sub-diffusive regime proportional to $t^\alpha$ driven by the action of colored noise, in agreement with the fLe's studied in Refs.~\cite{porra1996generalized,lutz2001fractional,tateishi2012different,sandev2014langevin}. This behavior holds when the linear friction term is neglected and the system goes out of equilibrium to keep both thermal baths. However, if the colored noise bath is neglected and the white noise causes the fractional friction, the dominant sub-diffusive regime grows as $t^{2-\alpha}$. The system reduces to the conventional Langevin equation for oBm exhibiting normal diffusion, when neglecting the colored noise bath and thermal gradient.

Periodic time-ordered phases emerge when the linear friction term is neglected, for small enough fractional orders  $\alpha\lesssim 0.1$. When the only contribution is given by the colored noise bath, the system exhibits the characteristics of a time crystal, with dominant ground state in thermal equilibrium and periodic time order in the MSD proportional to $2\pi$. On the other hand, when the only contribution is given by the white noise bath, fluctuation-dissipation is broken and the periodic time order of the MSD becomes proportional to $\pi$. A mixed time-ordered phase is found when the zero-linear friction system is coupled to both baths. In that case, the MSD shows a superposition of the ground and the non-equilibrium states with a dominant periodicity that deviates from $2\pi$ due to the superposition of the damping effects from each bath.

This research underscores, using both analytical and numerical methods, the periodic time-ordered phases emerging from the fLe with white and colored noise by investigating in-equilibrium and out-of-equilibrium regimes. Further explorations can be undertaken by extending the analytical and numerical approaches to more general correlated stochastic processes \cite{vinales2007anomalous, sandev2011generalized} and distributed-order equation systems \cite{ding2021applications}, to understand if other periodic time-ordered phases can emerge from the Caldeira-Leggett description and potentially diffuse at different scales.

\section{Acknowledgements}
We are grateful to P. Geraghty and C. Oosterlee for providing us with insights into the Monte Carlo method and the fBm generators, and to P. Guo for his feedback on the discretization of the fLe. We also thank M. Conte, D.F. Munoz-Arboleda, F.S. Abril-Bermúdez and E. Cantor for fruitful discussions. D.S.Q. acknowledges the financial support of the Ministry of Science Colombia (MinCiencias, Grant No. 906) and R.C.V. acknowledges the financial support of the Netherlands Organization for Scientific Research (NWO, Grant No. 680.92.18.05).

\appendix
\section{Review on Fractional Calculus}
\label{AppFC}

The definition of the Riemann-Liouville fractional derivative of order $m-1<\alpha<m$, with $m$ integer, is given by
\begin{equation}
    \prescript{RL}{a}{D}^{\alpha}_t f(t) = \frac{d^m}{dt^m} \int_a^t \frac{1}{\Gamma(m-\alpha)} (t-\tau)^{m-\alpha-1} f(\tau)d\tau.
\label{eq:FC_RL}
\end{equation}

Similarly, the definition of Caputo reads as
\begin{equation}
    \prescript{C}{a}{D}^{\alpha}_t f(t) = \int_a^t \frac{1}{\Gamma(m-\alpha)} (t-\tau)^{m-\alpha-1} f^{(m)}(\tau)d\tau.
\label{eq:FC_C}
\end{equation}

Note that both definitions are quite similar; however, the order in which integration and derivation are performed is opposite. This brings about important consequences in the definition of boundary conditions for physical problems. In particular, in the case of the fLe~\eqref{eq:fLe1}, it is more suitable to use the Caputo definition to avoid non-integer boundary conditions. However, when working with numerical discretizations, the Riemann-Liouville definition allows us to easily eliminate the derivative term, as shown in Section~\ref{sec:fLe-fdiff}.

By repeated integration by parts of the definition given by Eq.~\eqref{eq:FC_RL}, we can relate the Caputo and Riemann-Liouville derivatives by means of the following expression
\begin{equation}
    \prescript{C}{a}{D}^{\alpha}_t f(t) = \prescript{RL}{a}{D}^{\alpha}_t f(t) - \sum_{k=0}^{m-1} \frac{f^{(k)}(a)}{\Gamma(k-\alpha+1)}(t-a)^{k-\alpha},
\label{eq:FC-C_RL}
\end{equation}
where the upper indices $C$ and $RL$ refer to Caputo and Riemann-Liouville, respectively.

In particular, when $a = 0$, $m=0$ and $0<\alpha<1$, the previous expression reads
\begin{equation}
    \prescript{C}{0}{D}^{\alpha}_t f(t) = \prescript{RL}{0}{D}^{\alpha}_t f(t) - \frac{f(0)}{\Gamma(1-\alpha)}t^{-\alpha}.
\label{eq:FC-C_RL_0}
\end{equation}

\section{White and colored noise generators}
\label{AppNoise}
\begin{figure}[ht!]
    \centering
    \includegraphics[scale=0.45]{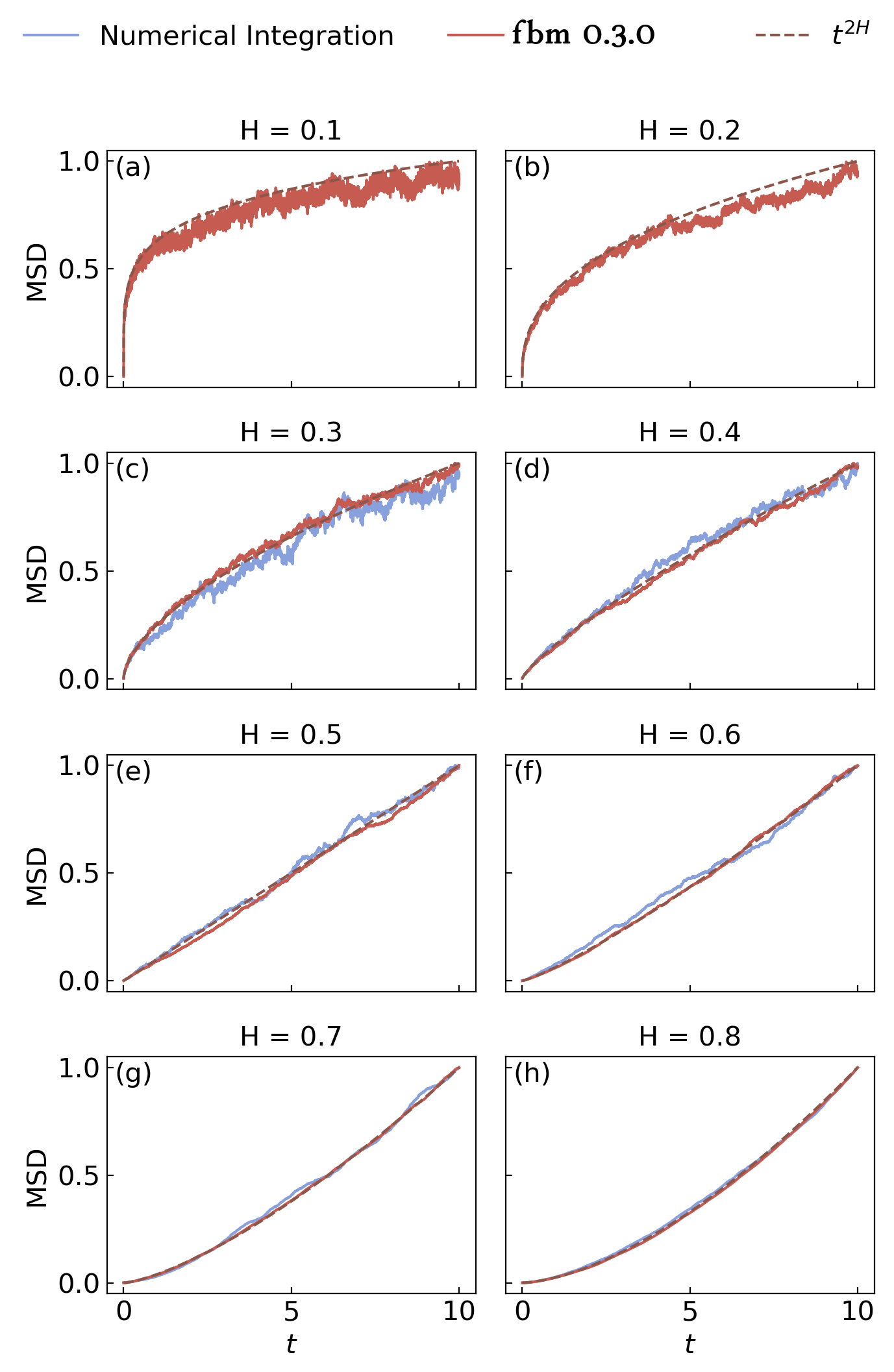}
    \caption{Mean squared displacement of fBm trajectories for different values of the Hurst exponent. The blue lines correspond to the numerical solution of the integral~\eqref{eq:BH_dis}, the red lines are the results of the package \textbf{fbm 0.3.0}, and the brown dashed lines are the analytical benchmark of the MSD, calculated with Eq.~\eqref{eq:corr_fBm}, assuming $t=\tau$.}
    \label{fig:BH-MSD}
\end{figure}

There exist several ways to simulate an fBm. The most direct solution is to numerically solve the integral~\eqref{eq:BH_man}, as in Refs.~\cite{decreusefond1999stochastic,geraghty2019financial}. Using the Euler hypergeometric function $\prescript{}{2}{F}_1$~\cite{askey1999special}, it is common to express this integral as
\begin{equation}
    B_{H} = \int_0^t K_{H}(t,\tau) dB(\tau),
\label{eq:BH_hyp}
\end{equation}
where
\begin{align}
    \begin{split}
        K_{H}&(t,\tau)= \\
        & \frac{(t-\tau)^{H-\frac{1}{2}}}{\Gamma(H+\frac{1}{2})}\;_2F_1\left (H-\frac{1}{2};\, \frac{1}{2}-H;\; H+\frac{1}{2};\, 1-\frac{t}{\tau} \right).
    \end{split}
\end{align}
To compute directly each step of the fBm, this equation can be discretized,
\begin{equation}
    B_{H}(t_j) = \frac{n}{T} \sum_{i=0}^{j-1} \Delta B_{i} \int_{t_i}^{t_{i+1}} K_{H}(t_j,\tau)d\tau,     
\label{eq:BH_dis}
\end{equation}
where $n$ is the number of steps between $t=0$ and $t=T$, and $\Delta B_{i}$ is an increment sampled from the uniform distribution $\sqrt{T}N(0,1)$. The integral can then be solved using any standard numerical method, such as the trapezoid rule or Gaussian quadrature.

The method described so far is computationally quite expensive due to the integration of hypergeometric functions. As an alternative, the Hosking \cite{hosking1984modeling} and the Davies-Harte~\cite{davies1987tests} methods can also be used to generate exact fBm trajectories. A glimpse of both methods can be found in Ref.~\cite{banna2019fractional}. The Hosking method relies on the Cholesky decomposition of the covariance matrix, which can make it heavy for long time series but more suitable for high values of $H$. In contrast, the Davies-Harte method efficiently generates fBm trajectories using FFT. In Fig.~\ref{fig:BH-MSD}, we compare the MSD of the simulated fBm using numerical integration of Eq.~\eqref{eq:BH_dis} and the implementation of the Davies-Harte and Hosking methods in the package of \textit{Exact methods for simulating fractional Brownian motion and fractional Gaussian noise in python} (\textbf{fbm 0.3.0})~\cite{flynn2019fbm}. We benchmark the results using Eq.~\eqref{eq:corr_fBm}, setting $t= \tau$. In particular, we observe that the numerical integration fails for $H\leq0.2$. Thus, we use Hosking method for high values of $H$ and Davies-Harte for the rest in order to speed up the numerical analysis, according to the prescription of the package.

\bibliography{bibliography}

\end{document}